\documentstyle[12pt,epsf]{article}
\begin{document}

\pagestyle{plain}

\renewcommand{\thefootnote}{\fnsymbol{footnote}}


\begin{flushright}
{\small
SLAC--PUB--8140\\
June 1999\\}
\end{flushright}

\begin{center}
{\large\bf
Supersymmetric Grand Unified Models without Adjoint Higgs Fields\footnote{Work supported by
Department of Energy contract  DE--AC03--76SF00515.}}

\smallskip

Chih-Lung Chou\\
Stanford Linear Accelerator Center, Stanford University, Stanford, CA 94309, USA\footnote{Address after August 1, 1999: Institute of Physics, Academia Sinica, Taipei 11529, Taiwan.}\\
\medskip

\end{center}

\vfill

\begin{center}
{\large\bf
Abstract }
\end{center}

We discuss two classes of supersymmetric grand unified theories based on extended gauge groups $SO(10) \times SO(10)$ and $SO(10)\times SO(10)\times SO(10)$.  Effective adjoint fields of each gauge group SO(10) are argued to be formed from combining two Higgs fields in fundamental representation of the extended gauge groups, one obtaining its VEV along the diagonal $SO(10)_D$ direction and the other acquiring its VEV along the diagonal $SU(5)_D\times U(1)_D$ or its subgroup direction. Thus experimentally acceptable fermion mass matrices, such as Georgi-Jarlskog ansatz, with successful GUT mass relations can be constructed in these theories.

\vfill

\section{Introduction}
The Standard Model (SM) provides a successful description of physics up to the weak scale.  However, it provides some 18 parameters which are input by hand to fit experiment data.  Most of these input parameters are associated with flavor physics and are included to parameterize the fermion mass hierarchy, Cabibbo-Kobayashi-Maskawa (CKM) angles, and neutrino oscillations.  Many theories, either supersymmetric (SUSY) or non-supersymmetric, are constructed to address the flavor problem and, hopefully, make predictions on new physics.  Among these theories beyond the Standard Model (SM), supersymmetric grand unification provides an elegant framework that explains not only the gauge quantum numbers of fermions transforming under the SM gauge group $SU(3)_C\times SU(2)_L\times U(1)_Y$, but also the prediction of $\alpha_s(M_Z)$. This remarkable success of the prediction of $\alpha_s(M_Z)$ motivates further exploration of SUSY grand unification \cite{SUSYunification}.

Among the ideas of grand unification, gauge groups such as $SU(5)$, $E_6$, and $SO(10)$ are frequently used in GUT model construction \cite{SUSYGUT,GUTbook}. However, there are reasons that make $SO(10)$ theories more attractive than others. First, $SO(10)$ is the smallest group in which all matter fields in one family can fit into one irreducible representation. Second, the two light Higgs doublets needed in any SUSY theory fit into one {\bf 10} of SO(10).  This allows the Yukawa couplings of up-type and down-type quarks to be determined by Clebsch-Gordan coefficients, thus making $SO(10)$ theories more predictive. 

There is a problem with this approach, however. Typical SUSY $SO(10)$ models need to use Higgs fields in higher representations, the {\bf 126} or {\bf 45}, to achieve successful GUT relations for Yukawa matrices. These representations are complex in their own right, and theories which contain tensor fields of rank higher than two cannot be constructed from the simplest string-derived GUT theories \cite{StringyGutReview,GGfromString}.  This motivates the use of extended GUT gauge groups such as $G\times G$ or $G\times G\times G$, where $G$ denotes the usual GUT group, in SUSY GUT model construction \cite{GG,GGG,Chou555}. 

 Supersymmetric GUT models based on the gauge groups $SO(10) \times SO(10)$ and $SO(10)\times SO(10)\times SO(10)$ have been discussed in the literature \cite{GG,GGG}. In these models, the breaking of the GUT gauge group was done when fundamental Higgs fields in the $(10, 10)$ representation, acquire their vacuum expectation values (VEVs) along the embedded diagonal subgroup directions of $SO(10)\times SO(10)$ and $SO(10)\times SO(10)\times SO(10)$, while the spinorial Higgs fields $\Psi_i, \bar \Psi_i$ acquire VEVs along $SU(5)$-preserving directions.  Four sets of the $(10, 10)$ fields carrying charges of different discrete symmetries were introduced; the large number of fields is needed not only to achieve the desirable Higgs doublet-triplet splitting, but also give the desirable asymmetry between the up  and down quark mass matrices. As a result, typical predictions of SUSY GUT $SO(10)$ models, such as the top-bottom Yukawa unification $\lambda_t = \lambda_b$, and Clebsch-Gordan relations in Yukawa matrices are not valid in their models.

In this paper, we follow the idea of using $SO(10)\times SO(10)$ and  $SO(10)\times SO(10)\times SO(10)$ as the SUSY GUT gauge groups. However, we show that the traditional merits of the SUSY GUT $SO(10)$ models can be preserved in our $SO(10)\times SO(10)$ and $SO(10)\times SO(10)\times SO(10)$ model construction. Although it is motivated from the string constructions, our model construction is self-contained and does not make explicit reference to string theories. In our models, all Higgs fields are in the fundamental representations of the gauge groups and no rank two tensors of any $SO(10)$ gauge group are required.  

In section \ref{sec:EffAd}, we show that the extended GUT gauge group breaking can be implemented when Higgs fields acquire VEVs along diagonal $SO(10)_D$ directions, diagonal $SU(5)_D\times U(1)_D$ directions, or other diagonal directions. Most importantly, we argue that the effective adjoint fields for each $SO(10)_i$ group can be formed by combining two VEV-acquiring Higgs fields.  In section \ref{sec:SUSY1010}, we construct an explicit model based on $SO(10)\times SO(10)$.  We show that the Higgs doublet-triplet problem is naturally solved through the Dimopoulos-Wilczek mechanism \cite{DWmechanism} without destabilizing the gauge hierarchy. The doublet-triplet splitting mechanism also guarantees strong suppression of proton decay, since the contributions from heavy Higgsino triplet exchange diagrams are absent or highly suppressed.  We also show that this model gives Yukawa matrices of the type similar to Georgi-Jarlskog ansatz. An explicit model which was analyzed by  Anderson {\it et al.} \cite{OperatorSO10} is constructed by using effective adjoint operators. In section \ref{sec:SUSY101010}, we present an $SO(10)\times SO(10)\times SO(10)$ model  with each family of matter multiplets transforming under different $SO(10)$ groups. In section \ref{sec:Conclusion10}, we make our conclusion.

\section{Effective adjoint operators for $SO(10)$}
\label{sec:EffAd}
As pointed out in the literature \cite{GG,GGG}, the breaking of extended GUT gauge groups $G\times G$ and $G \times G \times G$ can be achieved by a set of Higgs fields in the fundamental representation.  For example, an $SO(10)_1 \times SO(10)_2$ model breaks down to its diagonal subgroups when fields in the fundamental representation $(10, 10)$ develop VEVs. We will denote $(10, 10)$ fields in this paper as $S$ or $Z$ depending on the VEV patterns.  We denote fields with the following three canonical patterns of VEVs $<S_X>$, $<S_{B-L}>$, $<S_{T_{3R}}>$,  corresponding to

\begin{eqnarray}
<S_X>={v_D \over \sqrt{10}} \cdot {\scriptsize ( \begin{array}{cc}
				1&0\\
				0&1
				  \end{array} )}
\otimes \mbox{diag}(1,1,1,1,1) \\
<S_{B-L}>={v_G \over \sqrt{10}} \cdot {\scriptsize ( \begin{array}{cc}
				1&0\\
				0&1
				  \end{array} )}
\otimes \mbox{diag}(a,a,a,0,0)  \\
<S_{T_{3R}}>={v_G \over \sqrt{10}} \cdot{\scriptsize ( \begin{array}{cc}
				1&0\\
				0&1
				  \end{array} )}
\otimes \mbox{diag}(0,0,0,b,b).
\label{SVEVs}
\end{eqnarray}

\noindent The VEVs of  $S_X$, $S_{B-L}$, and $S_{T_{3R}}$ break $SO(10)_1\times SO(10)_2$ down to its embedded diagonal subgroups $SO(10)_D$, $SO(6)_D \times SO(4)_1 \times SO(4)_2$, and $SO(6)_1\times SO(6)_2\times SO(4)_D$ respectively. Usually, a tree level superpotential has many SUSY vacua which include the VEVs in Eq. (\ref{SVEVs}); a typical form includes   

\begin{equation}
W \supset {\lambda M \over 2}\mbox{Tr}(SS^T)+{A \over 4M}(\mbox{Tr}(SS^T))^2+{B \over 4M}\mbox{Tr}(SS^TSS^T).
\label{SVEVW}
\end{equation}

However, there are other SUSY vacua which lie along the direction of the embedded diagonal $SU(5)_D\times U(1)_D$, or other directions such as $SU(3)_D\times U(1)_D\times SO(4)_1\times SO(4)_2$ and $SO(6)_1\times SO(6)_2\times SU(2)_D\times U(1)_D$. We denote the associated $(10, 10)$ fields as $Z$ and again refer to the VEV patterns using subscripts:

\begin{eqnarray}
<Z_X>&=&{v_{10} \over \sqrt{10}} \cdot {\scriptsize ( \begin{array}{ccc}
				0&1\\
				-1&0
				  \end{array} )}
\otimes \mbox{diag}(2,2,2,2,2) \nonumber \\
<Z_{B-L}>&=&{v_5 \over \sqrt{10}} \cdot {\scriptsize ( \begin{array}{ccc}
				0&1\\
				-1&0
				  \end{array} )}
\otimes \mbox{diag}(2a/3,2a/3,2a/3,0,0) \nonumber \\
<Z_{T_{3R}}>&=&{v_5 \over \sqrt{10}} \cdot{\scriptsize ( \begin{array}{ccc}
				0&1\\
				-1&0
				  \end{array} )}
\otimes \mbox{diag}(0,0,0,b/2,b/2).
\label{ZVEVs}
\end{eqnarray}

As an alternative to generating scales or VEVs by minimizing a tree level superpotential, it has been shown that the scale $<S_X>$ could also be dynamically generated through a strongly coupled supersymmetric dynamics \cite{DynamicalScale}. Following the same line of thinking, we can introduce two supersymmetric gauge groups $SU(N_c)$ and $Sp(n_c)$ with fields in their fundamental representations $q(N_c,1,10,1)$, ${\bar q}({\bar N_c},1,1,10)$, and $Q(1,2n_c,1,10)$, where the numbers in brackets denote the dimensionality of each field under the two strong groups and the GUT gauge group $SO(10)_1 \times SO(10)_2$. With the imposition of some discrete symmetry, say $Z_N \times Z_K$, that keeps the field $S_X$ from coupling directly to $Z_X$, the lowest order of tree level superpotential is given by 

\begin{eqnarray}
W_{tree} = {A\over{2NM^{2N-3}}}S_X^{2N}+{B\over{2KM^{2K-3}}}Z_X^{2K}+\lambda_1 S_X q{\bar q}+{\lambda_2 \over M} (S_X Z_X)^{ab}Q^aQ^b \label{ZStree} 
\end{eqnarray}
\noindent  where $A$ and $B$ are coefficients, $M$ is the superheavy scale or Planck scale, and $\lambda_i$ denote the dimensionless coupling constants. To this should be added the dynamical superpotential resulting from the strong dynamics,
\begin{eqnarray}
W_{dyn}={C\over{N_c-10}}\left[ {\Lambda_1^{3N_c-10} \over \mbox{det}q{\bar q} } \right]^{1\over {N_c-10}}+{D \over {n_c+1-5}}\left[ {\Lambda_2^{3n_c -2} \over \mbox{Pf}(QQ) } \right]^{1 \over {n_c -4}}.
\label{StrongPotential}
\end{eqnarray}

\noindent By stabilizing the superpotential in Eqs (\ref{ZStree}) and (\ref{StrongPotential}) along the $<S_X>$ and $<Z_X>$ directions, we obtain the following equations for the VEVs:

\begin{eqnarray}
{A\over M^{2N-3}}s^{2N-1}+{5C \over (n_c+1) s}\left[{\lambda_2 sz\over M\Lambda_2}\right]^{5\over{n_c+1}}\Lambda_2^3+{10D \over N_c s}\left[{\lambda_1 s\over \Lambda_1}\right]^{10\over N_c}\Lambda_1^3=0 \label{StrongVEV1}\\
{B\over M^{2K-3}}z^{2K-1}+{5C \over (n_c+1) z}\left[{\lambda_2 sz\over M\Lambda_2}\right]^{5\over{n_c+1}}\Lambda_2^3=0, 
\label{StrongVEV2}
\end{eqnarray}

\noindent where $s=v_D/\sqrt{10}$ and $z=2v_{10}/\sqrt{10}$. It is easily seen that solving Eq.s (\ref{StrongVEV1}) and (\ref{StrongVEV2}) could lead to nonzero $v_{D}$ and $v_{10}$ when either one of the following conditions is satisfied:

\begin{eqnarray}
{2(n_c+1) \over N_c} &>& 1 + {10 \over N_c}, \hspace{0.5cm} \mbox{or} \\
{2(n_c+1) \over N_c} &\leq& 1.
\label{NonzeroVEV}
\end{eqnarray}

Given VEVs of the $S$ and $Z$ fields, we can form effective rank two tensors which carry quantum numbers of the gauge group $SO(10)_2$ by combining any two of the $S$ and $Z$ fields. In this way, we can form effective adjoint operators of $SO(10)_2$, which we call $\Sigma$ and $\Sigma '$, given by

\begin{eqnarray}
\Sigma^{bc}_X &\equiv& {1\over M}\mbox{Tr}_1(Z^T_XS_X)={1\over M}Z^{ab}_XS_X^{ac}, \nonumber\\
\Sigma^{bc}_{B-L}&\equiv& {1\over M}\mbox{Tr}_1(Z^T_{B-L}S_X)={1\over M}Z^{ab}_{B-L}S_X^{ac},  \nonumber \\
\Sigma^{bc}_{T_{3R}}&\equiv& {1\over M}\mbox{Tr}_1(Z^T_{T_{3R}}S_X)={1\over M}Z^{ab}_{T_{3R}}S_X^{ac}, \nonumber\\
{\Sigma'}^{bc}_{B-L} &\equiv& {1\over M}\mbox{Tr}_1(S^T_{B-L}Z_X)={1\over M}S^{ab}_{B-L}Z_X^{ac}\ \nonumber \\
{\Sigma'}^{bc}_{T_{3R}} &\equiv& {1\over M}\mbox{Tr}_1(S^T_{T_{3R}}Z_X)={1\over M}S^{ab}_{T_{3R}}Z_X^{ac}
\label{operator45}
\end{eqnarray}

\noindent We can also form effective identity operators of $SO(10)_2$, such as $I={1\over M}\mbox{Tr}_1(S_X^T S_X)$ or $I'={1\over M}\mbox{Tr}_1(Z_X^T Z_X)$.  Reciprocally, we can form effective adjoint and identity operators of $SO(10)_1$.  All of these effective tensors can arise physically from integrating out heavy states which transform under one of the $SO(10)$'s.  For example, we can generate the structure $\mbox{Tr}_1(Z^TZ')$ by integrating out the heavy states $10_1$ and $10_1 '$ from the following superpotential:

\begin{eqnarray}
M_110_110_1'+10_1Z10_2+10_1'Z'10_2' \longrightarrow {1\over M_1}10_2\mbox{Tr}_1(Z^TZ')10_2'
\label{OP45}
\end{eqnarray}

\noindent  Once we are equipped with these effective rank two tensors, it is possible to construct supersymmetric GUT models with realistic fermion masses and CKM angles.  A systematic analysis of the construction of $SO(10)$ GUT models has been done by Anderson {\it et al.} \cite{OperatorSO10}. Our treatment, with the $\Sigma$, $\Sigma'$, $I$ and $I'$ effective fields, now maps directly onto that analysis.

\section{A SUSY $SO(10)\times SO(10)$ GUT model}
\label{sec:SUSY1010}
In this section, we present an example based on the $S_X$ and $Z$ VEVs which demonstrates that typical SUSY $SO(10)$ GUT predictions can actually be preserved in $SO(10)_1 \times SO(10)_2$ gauge theories with experimentally acceptable Yukawa matrices. We assume four fundamental Higgs fields $S_X$, $Z_X$, $Z_{B-L}$, and $Z_{T_{3R}}$ of representation dimensionality $(10, 10)$ in our $SO(10)\times SO(10)$ GUT model. We construct the superpotential so that each of the $(10, 10)$ Higgs fields acquires a VEV along the indicated direction as described in Section \ref{sec:EffAd}.

\subsection{Higgs doublet-triplet splitting}
The Higgs structure is constructed by the requirement of Higgs doublet-triplet splitting. Higgs triplets, if they are not heavy enough, could contribute to the evolution of the gauge couplings, and thus spoil the unification of the gauge couplings.  In addition, Higgsino triplets may also mediate fast proton decay. So we might begin by analyzing the constraints imposed by the splitting mechanism.

In conventional $SO(10)$ models, Higgs triplet fields may acquire heavy masses by coupling to the adjoint fields which have their VEVs along the $B-L$ direction
\begin{eqnarray}
W(H_1, H_2)&=&H_1 A H_2, \hspace{1cm} \mbox{with} \nonumber \\
 <A>&=&V\cdot {\scriptsize ( \begin{array}{ccc}
				0&1\\
				-1&0
				  \end{array} )} \cdot \mbox{diag} (1,1,1,0,0),
\label{DWform}
\end{eqnarray}

\noindent where $H_1$ and $H_2$ are the fundamental Higgs fields, and $A$ denotes the adjoint Higgs field which acquires its VEV of the Dimopolous-Wilczek forms. As seen from Eq. (\ref{DWform}), the triplet fields in $H_1$ and $H_2$ get heavy masses $V$ and splitted from their doublet partners. 

In our $SO(10)\times SO(10)$ model, among the four fundamental Higgs, $Z_{B-L}$ and $Z_{T_{3R}}$ acquire their VEVs of the Dimopoulos-Wilczek (DW) forms  through the stabilization of a tree level superpotential as in Eq. (\ref{SVEVW}).  However, the DW forms of VEVs may be seriously destabilized when some cross coupling terms, such as $\mbox{Tr}(Z_{B-L}^T Z_{T_{3R}}) \equiv Z_{B-L}^{ab} Z_{T_{3R}}^{ab}$, $ \mbox{Tr}(Z_X^T Z_{B-L})$, $\mbox{Tr}(Z_X^T Z_{T_{3R}})$, and $\mbox{Tr}(Z_X^T Z_{B-L})$ are present in the superpotential.  For instance, the presence of the term $\mbox{Tr}(Z_X^T Z_{T_{3R}})$ would destabilizes the gauge hierarchy in $Z_{T_{3R}}$ since the F-flatness condition $\mbox{F}_{Z_{T_{3R}}}=0$ would give a term proportional to $Z_X$. As a result, these cross coupling terms must be excluded to implement the DW mechanism for the Higgs doublet-triplet splitting problem. Although SUSY allows unwanted superpotential terms to be dropped by hand, it is less arbitrary to forbid them by a discrete symmetry. 

Barr \cite{Stability55} has suggested that a discrete symmetry may do the job of forbidding the above cross coupling terms. In our model, there is a possible choice $K=Z_2^{T_{3R}} \times Z_2^{T_{3R}'} \times Z_2^{B-L}\times Z_5^1 \times Z_5^2$, and under which the various $Z$ fields transform as

\begin{eqnarray}
Z_2^{T_{3R}}&:& \hspace{0.6cm} Z_{T_{3R}} \rightarrow -Z_{T_{3R}} \nonumber \\
Z_2^{T_{3R}'}&:& \hspace{0.6cm} Z_{T_{3R}} \rightarrow -Z_{T_{3R}} \nonumber \\
Z_2^{B-L}&:& \hspace{0.6cm} Z_{B-L} \rightarrow -Z_{B-L} \nonumber \\
Z_5^1&:& \hspace{0.6cm} S_X \rightarrow e^{6 \pi i/5} S_X \nonumber \\
Z_5^2&:& \hspace{0.6cm} (S_X, Z_X) \rightarrow e^{2\pi i/5}(S_X, Z_X).
\label{Zsymmetry}
\end{eqnarray}

\noindent The $Z_2^{B-L}$ and $Z_2^{T_{3R}}$ symmetries in Eq. (\ref{Zsymmetry}) are designed to forbid the dangerous cross coupling superpotential terms noted above but still allow the coupling terms at the quartic level

\begin{eqnarray}
\mbox{Tr}(Z_{B-L}Z_{B-L}^T)\mbox{Tr}(Z_{T_{3R}} Z_{T_{3R}}^T)&,& \hspace{0.5cm} [\mbox{Tr}(Z_{B-L} Z_{T_{3R}}^T) ]^2 \nonumber \\
\mbox{Tr}(Z_{B-L}Z_{B-L}^T Z_{T_{3R}} Z_{T_{3R}}^T)&,& \hspace{0.5cm} \mbox{Tr}(Z_{B-L}Z_{T_{3R}}^T Z_{B-L}Z_{T_{3R}}^T).
\label{QuarticTerms}
\end{eqnarray}

\noindent The terms in Eq. (\ref{QuarticTerms}) might change the values of the scales appearing in $<Z_{B-L}>$ and $<Z_{T_{3R}}>$, but they do not destabilize the DW forms of VEVs.  The last two terms of Eq. (\ref{QuarticTerms}) would have zero contribution to the F-flatness conditions. However, they remove the would-be Goldstone modes that are not eaten by the gauge bosons in the fields $Z_{B-L}$ and $Z_{T_{3R}}$.  The $Z_2^{T_{3R}'}$ discrete symmetry prevents the effective identity operator $\mbox{Tr}_1(Z_X^T Z_{T_{3R}})$ from coupling to the spinorial superheavy states $\Psi_1$ and $\bar \Psi_7$ in our model.  However, this $Z_2^{T_{3R}'}$ symmetry is basically construction-dependent   and may not be necessarily introduced into our $SO(10)\times SO(10)$ model. We will come to this point again when discussing fermion spectrum in the next subsection.  In general, the symmetry $K$ would keep fields $S_X$ and $Z_X$ from coupling to $Z_{B-L}$ and $Z_{T_{3R}}$ up to a very high order, {\it e.g.} $\mbox{Tr}(Z_XZ_{B-L})\mbox{Tr}(Z_XZ_{B-L}S_X^8)$ as implied by Table \ref{tab:1010model}. Thus the DW forms of VEVs are protected up to corrections of the order of the weak scale when the GUT gauge group breaking parameters $v_D/M$, $v_{10}/M$ and $v_5/M$ are sufficiently small. 

An explicit superpotential giving Higgs doublet-triplet splitting by the above mechanism is:
 
\begin{equation}
W_{DT}=10_H Z_{B-L} 10_{H'}+{1\over M}10_{H'}S_XZ_{T_{3R}}10_{H''}+X10_{H''}10_{H''}
\label{DTpotential}
\end{equation}

\noindent where $10_H$, $10_{H'}$ and $10_{H''}$ denote Higgs fields in (1,10), (10,1), (10,1) representations respectively. $X$ is a gauge singlet\footnote%
{The singlet $X$ may or may not be the effective rank two tensor fields $I$ or $I'$ depending on how the $K$ symmetry is chosen in our model.}
 that acquires a GUT scale VEV and this makes $10_{H''}$ superheavy. The introduction of the singlet $X$ is required by the fact that if $10_{H''}10_{H''}$ is a singlet and present in superpotential, so is the non-renormalizable term $S_XS_X10_{H'}10_{H'}$.  This term $S_XS_X10_{H'}10_{H'}$, if exists, will give superheavy mass to the triplet states living in $10_{H'}$ and generates heavy Higgsino triplets exchange diagrams that mediate proton decay and spoil the strong suppression of proton decay. As in generating the effective rank two tensors, the non-normalizable term in Eq. (\ref{DTpotential}) may rise from integrating out heavy states in the (1, 10) representation. The insertion of the field $S_X$ in this term is designed to protect $10_{H}$ from coupling $10_{H''}$ to a high order level.  In order to achieve DW mechanism, these Higgs fields must transform non-trivially under the discrete symmetry $K$. In general, there are many possible ways of assigning $K$ charges to all fields in our model. One assignment for the $K$ charges is given and can be found in Table \ref{tab:1010model}  

 Here it is clear that the discrete symmetry $Z_5^1 \times Z_5^2$ would play the role of forbiding unwanted terms in superpotential.  According to Table \ref{tab:1010model}, the Higgs mass terms  $M_{HH}10_H10_H$ and $M_{H'H'}10_{H'}10_{H'}$ are forbidden by this discrete symmetry up to $(<X S_X^8>/M^8+<X^3S_X^2Z_X^2/M^6>)10_H10_H$ and $<S_X^2X^4>/M^{5}10_{H'}10_{H'}$ respectively, and $M_{HH''}10_H10_{H''}$ are very highly suppressed by the discrete symmetry $K$.  Therefore, up to the order of weak scale, the mass matrix $M_{H_T}$ and $M_{H_D}$ for Higgs triplets and doublets are given as

\begin{eqnarray}
M_{H_T}&=&\left({\small \begin{array}{ccc}
				0&<Z_{B-L}>&0\\
				<Z_{B-L}>&{<S_X^2X^4> \over M^5}&0\\
				0&0&<X>
				  \end{array}} \right), \nonumber \\
M_{H_D}&=&\left( {\small \begin{array}{ccc}
				0&0&0\\
				0&{<S_X^2X^4> \over M^5}&<{S_X Z_{T_{3R}}\over M}>\\
				0&<{S_X Z_{T_{3R}}\over M}>&<X>
				  \end{array} }\right).
\label{DTmatrices}
\end{eqnarray}

\noindent Therefore, as from Eq. (\ref{DTmatrices}), only one pair of the doublets in $10_{H}$ would remain light after the breaking of the GUT gauge group. However, 
as seen from Eq. (\ref{DTmatrices}), one pair of the heavy Higgs triplets may receive a GUT scale $M_G\sim v_5$ mass, while the corresponding Higgs doublets fields receive a mass $v_D^2 v_5^2/<X>$ which is less than the scale $v_5$. This may affect the gauge unification in our model depending upon the scale hierarchy between the two masses.  Actually, this mass discrepancy results from forbidding dangerous high order nonrenormalizable operators which also contribute to the mass matrices $M_{H_T}$ and $M_{H_D}$.  If we assume that only renormalizable terms in superpotential are allowed at the superheavy scale $M$ and all high order terms are generated from the Heavy Fermion Exchange mechanism (HFE) \cite{HFE}, than the scale ratio $<S_X>/M$ can be of order $O(1)$ thus all heavy Higgs states, doublets and triplets, would have GUT scale masses $m\sim v_5$ and we have the gauge unification as that of the minimal supersymmetric standard model (MSSM). In this paper, we simply assume negligible effects on gauge unification caused by the mass discrepancy among the heavy Higgs multiplets. 

Finally, we discuss the implications for proton decay. Eq. (\ref{DTmatrices}) also implies a strong suppression of proton decay. Since the high order operator $S_X^2X^410_{H'}10_{H'}/M^5 $ is present, then the dimension 5 operators (dimension 4 in superpotential) that mediate proton decay are formed by exchanging heavy Higgsino triplets  

\begin{equation}
{\lambda \over M^*}QQQL, \label{PdecayStrength},
\end{equation}

\noindent with the effective mass $M^*$

\begin{equation}
M^*\approx {M^5 <Z_{B-L}>^2 \over <S_X^2X^4>} \sim 10^{31} \mbox{ GeV} >> M_{pl}.
\label{PdecaySterngth}
\end{equation}

\noindent Here we use $Q$ and $L$ to represent the associated quarks and leptons in proton decay processes. The estimated value for $M^*$ in Eq. (\ref{PdecayStrength}) is obtained by assuming $M\sim M_{pl}$, $<Z_{B-L}>/<S_X> \sim 10^{-2}$, and $<X>/M\sim 10^{-4}$. The coupling strength parameter $\lambda \sim 10^{-7}$ comes from multiplying the associating Yukawa coupling constants in the color-Higgs exchange Feynman diagrams. To saturate current experiment limits on proton decay \cite{ParticleBook}, 
, the coupling strength $\lambda/M^*$ for the dimension 5 operators should be no large than about $10^{-24} \mbox{ GeV}^{-1}$.  Obviously, the estimated strength in Eq. (\ref{PdecayStrength}) is far more less than the limit, therefore proton decay is highly suppressed in our model. 

\subsection {Fermion masses}
Anderson {\it et al.} \cite{OperatorSO10} showed that, with adjoint operators $\Sigma$ in a SUSY GUT $SO(10)$ gauge theory, experimentally acceptable fermion mass spectrum as well as CKM angles can be obtained when these fields acquire their VEVs and break the GUT $SO(10)$ gauge group down to Standard Model gauge group. We can generate the same Yukawa matrices by using effective higher dimension operators.  These can be obtained by integrating out heavy fields.  Then, following the choices made by Anderson {\it et al.}, we show that viable fermion mass matrices, such as those incorporating  Georgi-Jarlskog ansatz \cite{OperatorSO10,Georgi}, can be constructed in the $SO(10)_1\times SO(10)_2$ model.

We need to introduce additional heavy fields in the {\bf 16} and ${\bf {\overline {16}}}$ of $SO(10)_2$.  We assume that all other matter multiplets also transform under the gauge group $SO(10)_2$. From Table \ref{tab:1010model} , it is easy to see that non-renormalizable terms at the quartic level, for instance the $\Psi_1 \mbox{Tr}_1(Z_X^T S_X) \bar\Psi_1$, are allowed to occur in our $SO(10)_1\times SO(10)_2$ model.  This term may come from integrating out a pair of superheavy spinorial fields $\Psi_1'(16,1)$ and $\bar\Psi_1'(\overline {16},1)$ from the renormalizable superpotential

\begin{equation}
W \supset M_1'\Psi_1' {\bar \Psi_1'}+\Psi_1S_X{\bar \Psi_1'}+\Psi_1' Z_X {\bar \Psi_1},
\label{3to4level}
\end{equation}

\noindent where $M_1'$ denotes the super-heavy mass of $\Psi_1'$ and $\bar\Psi_1'$. At the renormalizable level with the generated quartic terms, the most general tree level superpotential consistent with the discrete symmetry $K$ in Table \ref{tab:1010model} and responsible for giving masses to quarks and leptons has the form

\begin{eqnarray}
W_{mass} &\supset& 16_316_310_H+16_3\Sigma_{B-L}{\bar\Psi_2}+16_2{\Psi_1}10_H+16_2\Sigma_X{\bar\Psi_8}+16_1\Sigma_X{\bar\Psi_3}\nonumber \\
&+&\Psi_1\Sigma_X{\bar\Psi_1}+\Psi_2\Sigma_X{\bar\Psi_2}+\Psi_2\Sigma_{B-L}{\bar\Psi_1}+\Psi_3\Sigma_X{\bar\Psi_4}+\Psi_4\Sigma_X{\bar\Psi_5}\nonumber \\
&+&\Psi_5 \Psi_6 10_H+{\bar\Psi_6}\Sigma_X\Psi_7+{\bar\Psi_7}\Sigma_X\Psi_8+X_S16_2{\bar \Psi_2} \nonumber \\
&+&\sum^8_{i=3}\Psi_i \cdot I \cdot {\bar \Psi_i},
\label{WforMass}
\end{eqnarray}

\noindent where the gauge singlet field $X_S$ is introduced to give mass to $16_2$ and $\bar\Psi_2$ when acquiring a superheavy VEV. 
 
From Eq. (\ref{WforMass}), only the third family matter multiplet $16_3$ could  get a mass of weak scale due to the discrete symmetry $K$.  When the effective adjoint operators $\Sigma_X$ and $\Sigma_{B-L}$ acquire their VEVs, the spinorial fields $\Psi_i, \bar\Psi_i$ become heavy and can be integrated out in the low energy effective theory. The higher dimension operators $O_{ij}$ that give masses to matter quarks and leptons are thus generated after diagonalizing the mass matrices of these superheavy spinorial fields \cite{OperatorSO10}.

\begin{eqnarray}
O_{23}&=&16_2 10_H {\Sigma_{B-L}^2 \over \Sigma_X^2}16_3 \nonumber \\
O_{22}&=&16_2 10_H {X_S \Sigma_{B-L}\over \Sigma_X^2} 16_2 \nonumber \\
O_{12}&=&16_1 ({\Sigma_X \over I})^3 10_H ({\Sigma_X \over I})^3 16_2.
\label{SO10OPs}
\end{eqnarray}

\begin{figure}
\begin{center}
\leavevmode{\epsfxsize=6.00truein\epsfbox{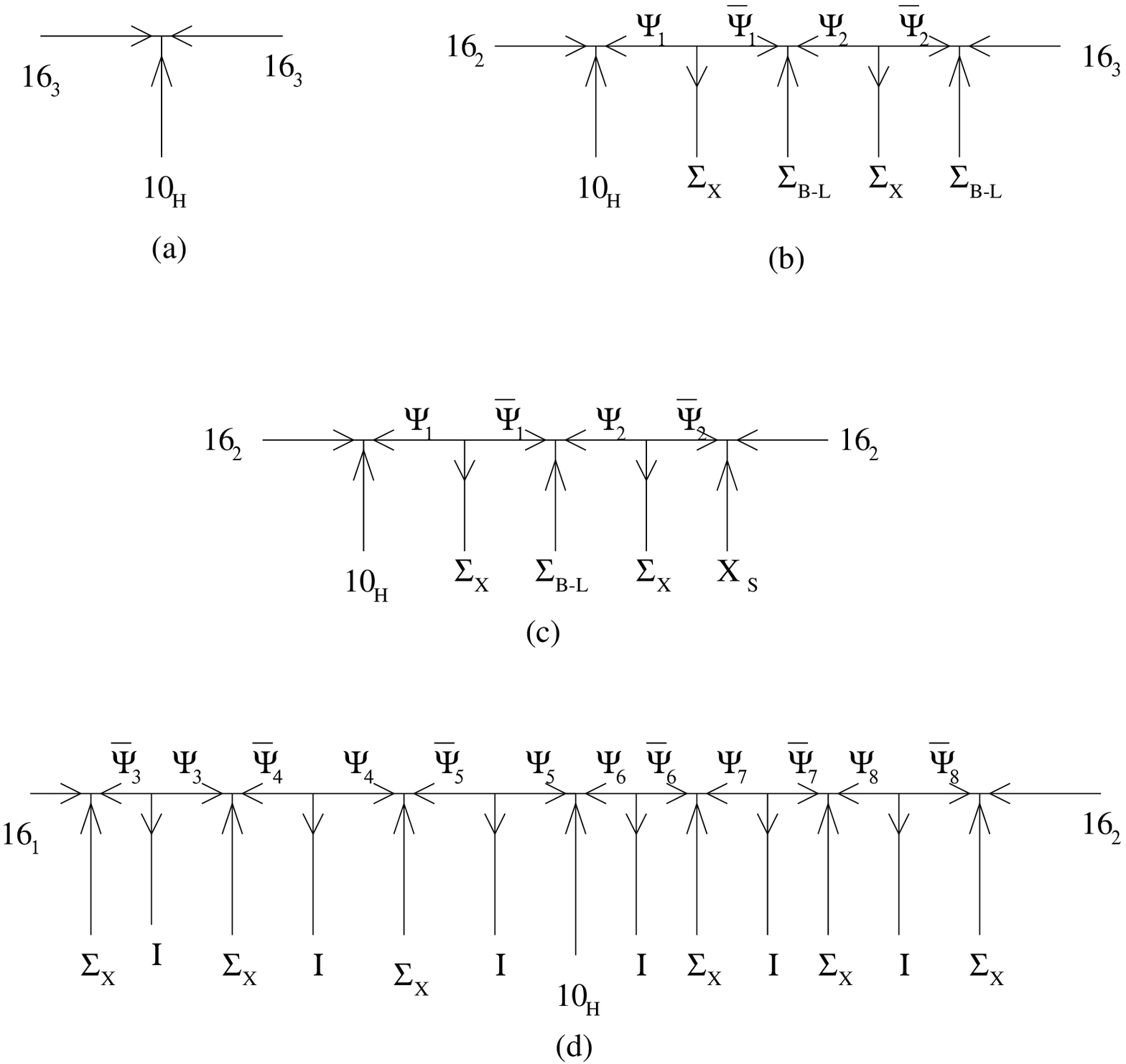}}
\end{center}
\caption{Operators $O_{ij}$ that give Yukawa matrices are formed by exchanging heavy fermion states.}
\label{fig:OijDiagram}
\end{figure}

\noindent The generation for the $O_{ij}$ operators is much easier to be seen from the diagrams in Fig.(\ref{fig:OijDiagram}). As seen from Eq. (\ref{SO10OPs}),  fermion mass hierarchy is explained due to the hierarchy of the GUT breaking scales $M>v_{D}>v_{10}>v_{5}$.  The effective adjoint operators $\Sigma_X$ and $\Sigma_{B-L}$ act on fermion states and give different quantum numbers to the states as described in Table \ref{tab:quantum45}.  As a result, Eq. (\ref{SO10OPs}) leads to the following typical Georgi-Jarlskog ansatz for the Yukawa matrices at GUT scale
\begin{eqnarray}
M_u&=&<H>\left[ \begin{array} {ccc}
		0&{1\over 27}C&0\\
		{1\over 27}C&0&B\\
		0&B&A
		\end{array} \right], \hspace{1cm} 
M_d=<\bar H>\left[ \begin{array} {ccc}
		0&-C&0\\
		-C&E&B\\
		0&{1\over 9}B&A
		\end{array} \right], \nonumber \\ 
M_e&=&<\bar H>\left[ \begin{array} {ccc}
		0&-C&0\\
		-C&3E&B\\
		0&9B&A
		\end{array} \right],
\label{GJMatrices}
\end{eqnarray}
\noindent  with $A\approx O(1)$, $B\approx v_5^2/v_{10}^2$, $C\approx 27(v_{10}^6/v_D^6)$, and $E\approx v_5M<X_S>/(v_D v_{10}^2)$. As seen from Eq. (\ref{GJMatrices}), we have the following successful GUT relations
\begin{eqnarray}
\lambda_t=\lambda_b&=&\lambda_{\tau} \label{taubottom}\\
\lambda_{22}^u:\lambda_{22}^d:\lambda_{22}^e&=&0:1:3 \label{lambda22}\\
m_{\tau}=m_{b}, \hspace{0.5cm}, m_{\mu}&\approx& 3m_s, \hspace{0.5cm} m_d\approx 3m_e, \label{GJrelations}
\end{eqnarray}

\noindent where $\lambda$'s denote the effective Yukawa coupling constants for corresponding mass operators.

Conclusively, it is suggested that the breaking of our GUT model is arranged as 

\begin{eqnarray}
SO(10)_1\times SO(10)_2 &{\stackrel  {v_D}{\longrightarrow}}& SO(10)_D \nonumber \\
&{\stackrel {v_{10}}{\longrightarrow}}& SU(5)_D\times U(1)_D \nonumber \\
&{\stackrel {v_5}{\longrightarrow} }& SU(3)_C\times SU(2)_L\times U(1)_Y,
\label{breaking1}
\end{eqnarray}

\noindent with approximate ratios $v_D/M \sim 1/30$, $v_{10}/v_D\sim O(10^{-1})$, $v_5/v_{10}\sim O(10^{-1})$, and $<X_S>=<X>\approx v_Dv_5/M$.  Detailed analysis for the mass operators $O_{ij}$ can be found in \cite{OperatorSO10}, and will not be discussed in this paper.  

As in more familiar GUT $SO(10)$ models, we can also analyze the neutrino masses in our $SO(10)_1\times SO(10)_2$ model.  First we observe that the matrix $M_{\nu^c \nu}$ for Dirac masses of neutrinos has a nonzero 22 entity also coming from $O_{22}$, and is far from identical to the up-quark mass matrix.    

\begin{equation}
M_{\nu^c \nu}=<H>\cdot \left[
		\begin{array}{ccc}
		0&-125C&0 \\
	      -125C&-{6 \over 25} E&B\\
		0&{9\over 25}B&A
		\end{array} \right]
\label{Mneu}
\end{equation} 

\noindent Since $125C$ is almost the same order of magnitude as $A$, the Dirac mass matrix for neutrino is no longer as hierarchical as quark and charged lepton mass matrices.  To form Majorana mass for the right handed neutrinos, we introduce a set of spinorial Higgs fields $\Psi_{V_i}(1,16)$, $\bar \Psi_{V_i}(1, \overline {16})$ which VEVs preserve the $SU(5)_2$ subgroup of $SO(10)_2$.  In general, the following neutrino mass operators can also be formed from heavy fermion exchanges

\begin{equation}
{1\over M}\sum_{i,j}(\bar \Psi_{V_i} \Gamma_a^{\over {(126)}} \bar \Psi_{V_j})(16_i \Gamma_a^{(126)} 16_j),
\label{MR}
\end{equation}
\noindent where $i, j$ are flavor indices. For simplicity, we would assume the Majorana mass matrix $M_R$ for right handed neutrinos to be a diagonal matrix with eigenmasses $m_{R_1}$, $m_{R_2}$ and $m_{R_3}$. Thus, from Eqs (\ref{Mneu}) and (\ref{MR}), the effective left handed Majorana mass matrix is

\begin{equation}
M_{\nu \nu}\approx M_{\nu^c \nu}^{+} M_R^{-1} M_{\nu^c \nu}.
\label{ML}
\end{equation}

\noindent Taking $C/E \approx 0.22 \approx 6/25$ as implied by the Cabibbo angle, it thus lead to the following three Majorana eigenmasses for left handed neutrinos 

\begin{equation}
m_{\nu_{\tau}}\approx {m_t^2 \over m_{R_3}}, \hspace{0.6cm} m_{\nu_{\mu}}\approx {(125 m_d tan\beta)^2 \over m_{R_2}}, \hspace{0.6cm} m_{\nu_{e}}\approx {(125 m_d tan\beta)^2 \over m_{R_1}}.
\label{MLeigenvals}
\end{equation}

\noindent Although all the VEVs of $\Psi_{V_i}$ need not to be the same,  we might take all $<\Psi_{V_i}>$ to be equal to $v_{10}$ for illustration. This gives $m_{R_i}=m_R\approx 2\times 10^{14} \mbox{ GeV}$ and leads to $m_{\nu_{\tau}} \sim 1/20 \mbox{ eV}$, $m_{\nu_u} \approx m_{\nu_e} \sim O(10^{-3} \mbox{ eV})$ for $\mbox{tan}\beta=45$.  Visible ${\nu_{\tau}}-{\nu_{\mu}}$  and ${\nu_{\tau}}-{\nu_{e}}$ oscillations with very small  neutrino oscillation angles $\mbox{sin}^22\theta_{\tau \mu}^{osc}$ and $\mbox{sin}^22\theta_{\tau e}^{osc}$ are favored when taking such assumption. However, other mass spectra for left handed neutrinos as well as large neutrino mixing angles may be obtained \cite {SO10neutrino,U1xFlavor}. This is because the right handed neutrino mass matrix $M_R$ may itself be nontrivial and have a hierarchical
 structure in our $SO(10)\times SO(10)$ model. 

\section{An SO(10) $\times$ SO(10) $\times$ SO(10) model}
\label{sec:SUSY101010}
It is straightforward to extend the GUT gauge group to $SO(10)_1\times SO(10)_2\times SO(10)_3$ and have all matter multiplets transform under one of the $SO(10)$ gauge groups.  However, this extension is basically a replication of the SUSY GUT $SO(10)_1\times SO(10)_2$ model described in the previous sections.  Different from the above direct generalization, in this section, we assign each matter multiplet $16_i$ to transform under different gauge group $SO(10)_i$.  We also assume the existence of the three Higgs fields $10_H(1,1,10)$, $10_{H'}(1,10,1)$, and $10_{H''}(1,10,1)$, and a set of fundamental Higgs fields $S_X(1,10,10)$, $Z_X(1,10,10)$, $Z_{B-L}(1,10,10)$, $Z_{T_{3R}}(1,10,10)$, $Z_{B-L}'(10,10,1)$, and $Z_{T_{3R}}'(10,10,1)$ for implementing the DW mechanism. The complete set of assignment is shown in Table \ref{tab:103fields}. 

The fundamental Higgs fields acquire their VEVs along some GUT breaking directions as described in the previous sections.  To protect the DW forms of the VEVS, some discrete symmetries above the GUT scales must typically be expected to restrict possible tree level superpotential terms. Without giving the discrete symmetries explicitly, we note that the superpotential responsible for giving heavy masses to Higgs triplet fields must be restricted to the following form:

\begin{eqnarray}
10_H Z_{B-L} 10_{H'}+{1\over M}10_{H'}S_X Z_{T_{3R}}10_{H''}+X10_{H''}10_{H''}, 
\label{DWGGG}
\end{eqnarray} 

\noindent where $M$ denotes the superheavy scale and $X$ is again a gauge singlet with a GUT scale VEV.  By the HFE mechanism mentioned in Section \ref{sec:SUSY1010}, the second term in Eq. (\ref{DWGGG}) may also come from integrating out some superheavy states.  In the worst case, if all allowed nonrenormalizable operators are present in superpotential, the gauge hierarchy as well as the DW forms of VEVs could still be protected up to a very high order by some discrete symmetries. As a result, only the pair of the Higgs doublets states in $10_H$ remain light down to the weak scale and proton decay could be suppressed strongly.  

In the following, we will briefly discuss the construction of realistic fermion mass matrices without going into details of how the fields transform under the needed discrete symmetries. 

As usual, only the third family of matter multiplet $16_3$ gets a weak scale mass through the tree level dimension four operator $O_{33}=16_316_310_H$.  Other $O_{ij}$ operators are generically nonrenormalizable because $16_1$ and $16_2$ both carry no $SO(10)_3$ gauge quantum numbers.  However, it is impossible to form $O_{ij}$ operators for the off-diagonal entries of fermion mass matrices by simply using matter multiplets and the Higgs fields in fundamental representations. A set of additional heavy fields in the ${\bf 16+{\overline {16}}}$, $\Psi_{V_i}$ and $\bar \Psi_{V_i}$, which transform under the GUT gauge group as $(16,1,1)$, $({\overline {16}},1,1)$, $(1, 16, 1)$, $(1, {\overline {16}}, 1)$, $(1,1,16)$ and $(1,1,{\overline {16}})$, must be introduced into the model and acquire VEVs along the $SU(5)_i$ singlet directions. The VEV's can be stabilized by the superpotential of the following form \cite{linearW}
\begin{equation}
Y(\Psi_V {\bar \Psi_V})^2/M_V^2+ f(Y),
\label{VEV16}
\end{equation}
\noindent where Y is a singlet field and $f(Y)$ is a polynomial function that contains a linear term. Notice that the would-be Goldstone modes in the spinors $\Psi_{V_i}$ and $\bar \Psi_{V_i}$ can be removed by adding more terms to the superpotential \cite{linearW}. 

Although there are many possible nonrenormalizable operators which may or may not survive from imposing the discrete symmetries, the following high dimensional operators could also arise from the HFE mechanism, and are interesting because they may help to realize the Gerogi-Jarlskog type of Yukawa matrices in the model.

\begin{eqnarray}
O_{23}&=&(\Psi_{V_2} \cdot S_X \cdot S_X \cdot 16_2) \cdot (\Psi_3 \cdot 10_H \cdot {1 \over \Sigma_X^2}16_3)\\ 
O_{22}&=&(16_2 \cdot S_X \cdot {\Sigma_{B-L} \over \Sigma_X^2}16_2) \cdot 10_H \\
O_{12}^{(1)}&=&(\Psi_{V_1} \cdot Z_{B-L}' \cdot 16_1) \cdot (\Psi_{V_2} \cdot S_X \cdot 16_2) \cdot 10_H \\
O_{12}^{(2)}&=&(\Psi_{V_1}' \cdot Z_{T_{3R}}' \cdot 16_1) \cdot (\Psi_{V_2} \cdot S_X \cdot 16_2) \cdot 10_H \\
O_{12}^{(3)}&=&(\Psi_{V_1} \cdot Z_{B-L}'\cdot Z_{B-L}' \cdot 16_1) \cdot (\Psi_{V_2}' \cdot S_X \cdot {\Sigma_X^3 \over I^3} 16_2) \cdot 10_H \\
O_{12}^{(4)}&=&(\Psi_{V_1} \cdot Z_{T_{3R}}' \cdot Z_{T_{3R}}' \cdot 16_1) \cdot (\Psi_{V_2}' \cdot S_X \cdot {\Sigma_X^3 \over I^3}16_2) \cdot 10_H \\
O_{12}^{(5)}&=&(\Psi_{V_1}' \cdot Z_{B-L}'\cdot Z_{T_{3R}}' \cdot 16_1) \cdot (\Psi_{V_2}' \cdot S_X \cdot {\Sigma_X^3 \over I^3} 16_2) \cdot 10_H 
\label{OPGGG}
\end{eqnarray}

\noindent Again, the effective adjoint operator $\Sigma_X$ of the gauge group $SO(10)_2$ gives different quantum numbers to the fermion states in the matter multiplets $16_i$.  Since the Higgs fields $Z'_{B-L}$ and $Z'_{T_{3R}}$ must at least carry different charges of some global $Z_2$ symmetry to avoid the breaking of gauge hierarchy, we thus need two additional VEV-acquiring spinors ${\Psi'}_{V_{i}}$ and $\overline {\Psi'}_{V_{i}}$ , where $i=1,2$, to make the operators $O_{12}$ respect the $Z_2$ symmetry. 

Let us parametrize the contributions of the operators $16_3 16_3 10_H$ and the $O_{ij}$ as $A, B, E,....$. In the case that only $16_3 16_3 10_H$ and the $O_{ij}$ operators give dominant contributions to fermion masses, the fermion mass matrices become

\begin{eqnarray}
M_u&=&<H>\left[ \begin{array} {ccc}
		0&C^{(3)}&0\\
		C^{(5)}&0&B\\
		0&B&A
		\end{array} \right], \nonumber \\
M_d&=&<\bar H>\left[ \begin{array} {ccc}
		0&C^{(1)}&0\\
		27C^{(5)}&E&0\\
		0&{1\over 9}B&A
		\end{array} \right], \nonumber \\ 
M_e&=&<\bar H>\left[ \begin{array} {ccc}
		0&27C^{(4)}&0\\
		C^{(2)}&3E&{1\over 9}B\\
		0&0&A
		\end{array} \right], \nonumber \\
M_{\nu^c \nu}&=&<H>\cdot \left[
		\begin{array}{ccc}
		0&27(C^{(3)}+C^{(4)}+C^{(5)})&0 \\
	      C^{(2)}&-{6 \over 25} E&{1\over 9}B\\
		0&0&A
		\end{array} \right],
\label{MassGGG}
\end{eqnarray}

\noindent where $A$, $B$, $E$, and $C^{(i)}$ come from the contribution of the operators $O_{33}$, $O_{23}$, $O_{22}$, and $O_{12}^{(i)}$ respectively.  Again, the numbers shown in Eq. (\ref{MassGGG}) are Clebsch-Gordan coefficients. From the mass ratio $m_u/m_d$,  we may estimate that the ratio $C^{(3)}/C^{(1)} \approx 1/27$.  Therefore, to realize an experimentally acceptable fermion mass matrix, as implied from Eq. (\ref{MassGGG}), the breakdown of the GUT gauge group $SO(10)_1 \times SO(10)_2 \times SO(10)_3$ may take the following steps

\begin{eqnarray}
[SO(10)]^3 &\longrightarrow& SU(5)_1 \times SU(5)_2 \times SU(5)_3 \hspace{0.6cm} \mbox{at} <\Psi_{V_i}> \sim M \nonumber \\
&\longrightarrow& SU(3)_C \times SU(2)_L \times U(1)_Y \hspace{0.6cm}  \mbox{at} \hspace{0.3cm} v_5 \approx {1\over 30}M,
\end{eqnarray}

\noindent where $M\approx 6\times 10^{17} \mbox{ GeV}$, $v_D \approx v_{10}\approx v_5$, and $<X_S>/v_5 \sim <X_S'>/v_5 \sim 10^{-1}$ are assumed. In this GUT group breaking scenario, the $SO(10)\times SO(10)\times SO(10)$ GUT gauge group would first be broken down to $SU(5)\times SU(5)\times SU(5)$ by the spinorial Higgs fields $\Psi_{V_i}$ and $\bar\Psi_{V_i}$, and then breaks into the embedded diagonal subgroup $SU(3)_C \times SU(2)_L\times U(1)_Y$ at the GUT scale $M_G \approx v_5$.

Neutrinos may acquire masses by the same mechanism described in the previous section.  A set of spinorial Higgs fields with non-vanishing VEVs along the $SU(5)_i$-preserving directions are necessary for giving Majorana masses to right handed neutrinos.  However, none of the spinorial Higgs fields used in constructing the operators $O_{ij}$ can be used in giving a Majorana mass to right handed $\tau$ neutrino $\nu_{\tau}$ since, otherwise, we would get the  Majorana mass for left handed $\tau$ neutrino $m_{\nu_{\tau}} \approx m_t^2/M \approx 1/6 \times 10^{-4} \mbox{eV}\gg m_{\nu_{\mu}}, m_{\tau_{e}}$, which is disfavored by recent SuperKamiokande data \cite{SuperK}. Thus a new pair of spinorial Higgs fields $\Psi_{V_3}'$ and $\Psi_{V_3}'$ would be needed to give an acceptable mass to $\nu_{\tau}^c \nu_{\tau}^c$

\begin{eqnarray}
{1\over M} (\bar \Psi_{V_3}' \Gamma_a^{\over {(126)}} \bar \Psi_{V_3}')(16_3 \Gamma_a^{(126)} 16_3).	
\label{neutauR}
\end{eqnarray}

\noindent with $<\Psi_{V_3}'>=<\bar \Psi_{V_3}'>\approx v_5$.

As before, a non-trivial Majorana mass matrix $M^R_{\nu}$ for right handed neutrinos may be present in the model, and heavily influence the Majorana mass spectrum as well as neutrino mixing angles of left handed neutrinos. We will not discuss this problem in detail in this paper.

\section{Conclusion}
\label{sec:Conclusion10}
Typical SUSY $SO(10)$ GUT models require a variety of rank two tensor fields, such as the fields in ${\bf 45}$ and ${\bf 54}$ representations, to be phenomenologically successful.  These rank two tensors, when they acquire their VEVs and break the gauge $SO(10)$ group, also play important roles in implementing the Dimopoulos-Wilczek mechanism and in deriving experimentally acceptable Yukawa matrices.  However, these representations are complicated, and it is usually difficult for all the needed rank two tensor fields to be generated by a simple string construction. 

In this paper, we have shown that it is possible not only to implement the DW mechanism but also to provide experimentally acceptable Yukawa matrices. In our $SO(10)\times SO(10)$ and $SO(10)\times SO(10)\times SO(10)$ GUT models, without introducing any rank two tensor fields, the Higgs doublet-triplet splitting problem is naturally solved with strong suppression of proton decay when some Higgs fields of fundamental representations acquiring their VEVs in Dimopoulos-Wilczek forms.  Also, unlike other $SO(10)\times SO(10)$ and $SO(10)\times SO(10)\times SO(10)$ models in the literatures \cite{GG,GGG}, effective adjoint operators of at least one of the $SO(10)$ gauge group can be formed when combining the $S$ and one of the $Z$ fields in our model.  This allows us to construct realistic fermion mass matrices with successful GUT relations such as top-bottom-tau unification $\lambda_t = \lambda_b =\lambda_{\tau}$, $m_{\mu}=3m_s$, and $m_d=3m_e$.  

On the neutrino mass problem, as in conventional $SO(10)$ theories, some spinorial Higgs fields in the {\bf 16} representation of the corresponding $SO(10)_i$ gauge group are necessary for making effective $\nu^c \nu^c$ mass operators.  When acquiring VEVs that preserve subgroups $SU(5)_i$ for each corresponding $SO(10)_i$, these spinorial Higgs fields give superheavy Majorana masses to right handed neutrinos.  Small Majorana masses for left handed neutrinos are thus generated from see-saw mechanism.  However, further understandings on the neutrino sector will be needed in our models for constructing the mass matrix for right handed neutrinos, and also for understanding the mass hierarchy/splitting as well as the mixing angles among left handed neutrinos.

\subsection*{Acknowledgments}

The author thanks Michael E. Peskin for very helpful comments. This work was supported by Department of Energy under contract DE-AC03-76SF00515.


\newpage
\newpage

\begin{table}
\begin{center}
\begin{tabular}{|c|c|c|c|c|c|}
\hline 
$\Psi_1$&$\Psi_2$&$\Psi_3$&$\Psi_4$&$\Psi_5$&$\Psi_6$\\
\hline
(+,+,2,-2)&(-,-,2,-1)&(-,+,2,-2)&(+,+,-1,-2)&(-,+,1,-2)&(-,+,-2,2)\\
\hline
$\bar\Psi_1$&$\bar\Psi_2$&$\bar\Psi_3$&$\bar\Psi_4$&$\bar\Psi_5$&$\bar\Psi_6$\\
\hline
(-,+,0,0)&(+,-,0,-1)&(-,+,2,0)&(+,+,0,0)&(-,+,-2,0)&(-,+,1,1)\\
\hline
$\Psi_7$&$\Psi_8$&$S_X$&$Z_X$&$Z_{B-L}$&$Z_{T_{3R}}$\\
\hline
(+,+,1,2)&(-,+,-1,2)&(+,+,3,1)&(-,+,0,1)&(+,-,0,0)&(-,+,0,0)\\
\hline
$\bar\Psi_7$&$\bar\Psi_8$&$X_S$&$16_1$&$16_2$&$16_3$\\
\hline
(+,+,-2,1)&(-,+,0,1)&(+,-,-2,-1)&(+,+,-2,0)&(+,+,2,2)&(+,+,2,0)\\
\hline
$X$&$10_H$&$10_{H'}$&$10_{H''}$&&\\
\hline
(+,+,-1,2)&(+,+,1,0)&(+,-,-1,0)&(-,-,-2,-1)&& \\ \hline
\end{tabular}
\end{center}
\caption{Fields transforming under the discrete symmetry $Z_2^{T_{3R}} \times Z_2^{B-L} \times Z_5^1 \times Z_5^2$. All fields are $Z^{T_{3R}'}_2$ singlets except for the fields $Z_{T_{3R}}$ and $10_{H''}$.}
\label{tab:1010model}
\end{table}

\newpage
\begin{table}
\begin{center}
\begin{tabular}{rrrrr}
\hline
\hspace{2cm}&\hspace{1.8cm}$X$&\hspace{1.6cm}$B-L$&\hspace{1.6cm} $T_{3R}$ &\hspace{1.6cm}Y\\
\hline
$u$&1&1&0&1/3\\
$\bar u$&1&-1&-1/2&-4/3\\
$d$&1&1&0&1/3\\
$\bar d$&-3&-1&-1/2&2/3\\
$e$&-3&-3&0&-1\\
$\bar e$&1&3&1/2&2\\
$\nu$&-3&-3&0&-1\\
$\bar \nu$&5&3&-1/2&0\\
\hline
\end{tabular}
\end{center}
\caption{Quantum numbers of the adjoint {\bf 45} VEVs on fermion states.}
\label{tab:quantum45}
\end{table}

\newpage

\begin{table}[t]
\begin{center}
\begin{tabular}{|cccc|}
\hline
  		& \hspace{0.5cm}$SO(10)_1$ \hspace{0.5cm}& \hspace{0.5cm}$SO(10)_2$ \hspace{0.5cm}& \hspace{0.5cm}$SO(10)_3$ \hspace{0.5cm} \\ \hline
$S_X$	&          	& 10            & 10 \\
$Z_X$   & 	        & 10     & 10 \\
$Z_{B-L}$ 		& 		& 10		& 10 \\
$Z_{T_{3R}}$		& 	& 10 	& 10 \\
${Z'}_{B-L}$ 		& 10		& 10	&   \\ 
${Z'}_{T_{3R}}$		& 10		& 10		& \\ \hline
$10_H$		&  		& 		& 10	\\
$10_H'$		& 		&  10		&  \\
$10_H''$	&		& 	10	& \\
$16_1$& 16 &&\\
$16_2$& & 16 &\\
$16_3$&&&16 \\
\hline
\end{tabular}
\end{center}
\caption{The field content of the $SO(10)_1\times SO(10)_2\times SO(10)_3$ model.}
\label{tab:103fields}
\end{table}
\end{document}